\documentclass[pra,aps,twocolumn,showpacs,superscriptaddress,nofootinbib]{revtex4}
\usepackage{graphicx}
\usepackage{dcolumn}
\usepackage{bm}
\usepackage{amsmath}
\usepackage{epsfig}
\usepackage{hyperref}
\usepackage[all]{hypcap}

\def\ii{{\rm i}}  \def\ee{{\rm e}}  
\def\Rb{{\bf R}}      
  \def\kb{{\bf k}}  \def\Eb{{\vec{\mathcal{E}}}}  \def\rb{{\bf r}}
    \def\zz{\hat{\bf z}}    \def\rr{\hat{\bf r}}

  \def\me{{m_{\rm e}}}  \def\lll{{\ell}}
\def\kperb{{{\bf k}_\perp}}  
\def\lambdap{{\lambda_{\rm p}}}

\usepackage{color}

\begin{document}
\title{Electron Diffraction by Plasmon Waves}

\author{F.~J.~Garc\'{\i}a~de~Abajo}
\email[Corresponding author: ]{javier.garciadeabajo@icfo.es}
\affiliation{ICFO-Institut de Ciencies Fotoniques, The Barcelona Institute of Science and Technology, 08860 Castelldefels (Barcelona), Spain}
\affiliation{ICREA - Instituci\'o Catalana de Recerca i Estudis Avan\c{c}ats, Barcelona, Spain}
\author{B.~Barwick}
\affiliation{Department of Physics, Trinity College, Hartford, Connecticut 06106, United States}
\author{F.~Carbone}
\affiliation{Laboratory for Ultrafast Microscopy and Electron Scattering, ICMP, \'Ecole Polytechnique F\'ed\'erale de Lausanne, Station 6, CH-1015 Lausanne, Switzerland}
\date{\today}

\begin{abstract}
An electron beam traversing a structured plasmonic field is shown to undergo diffraction with characteristic angular patterns of both elastic and inelastic outgoing electron components. In particular, a plasmonic {\it grating} (e.g., a standing wave formed by two counter-propagating plasmons in a thin film) produces diffraction orders of the same parity as the net number of exchanged plasmons. Large diffracted beam fractions are predicted to occur for realistic plasmon intensities in attainable geometries due to a combination of phase and amplitude changes locally imprinted on the passing electron wave. Our study opens new vistas in the study of multiphoton exchanges between electron beams and evanescent optical fields with unexplored effects related to the transversal component of the electron wave function.
\end{abstract}
\maketitle


\section{Introduction}

The weakness of free-space electron-photon interactions at low photon energies ($\ll\me c^2$) is clearly emphasized by the correspondingly small Thomson scattering cross-section $(8\pi/3)(\alpha\hbar/\me c)^2\sim10^{-28}\,$m$^2$. It is thus understandable that nearly seven decades went by between the proposal of electron diffraction by free light beams \cite{KD1933} (the so-called Kapitza-Dirac effect) and its demonstration in a set of elegant experiments by Batelaan and coworkers \cite{FAB01,FB02,B07}. Periodic motion of the electron aimed by the light electric field leads to a nonzero-average transversal Lorentz force due to the magnetic field, which causes elastic deflection of the electron, and consequently imprints spatial phase variations on the electron wave function.

Significant electron-photon coupling can occur when light is slowed down in dielectric media, essentially circumventing the large kinematic mismatch between electrons and photons at nonrelativistic energies. The resulting Cherenkov radiation \cite{C1934,FT1937,G96} has been instrumental in the development of particle detectors, whereas the already demonstrated inverse Cherenkov effect \cite{EKP1981} holds potential for electron acceleration \cite{KKR95}. While these phenomena are impractical for moderate electron energies in homogeneous materials, a related effect associated with the coupling to optical modes in photonic-crystals has been demonstrated for electrons moving in vacuum through holes perforating thin films \cite{paper048}.

Evanescent light fields offer an efficient way to enhance electron-photon interaction. This is the principle underlying the Smith-Purcell effect \cite{SP1953}, which in its inverse form \cite{MPN1987} also involves electron deflection by light fields. Plasmons sustained by nanostructured conductors (a prototypical class of evanescent fields), can extend toward the surrounding vacuum and enable large interaction with free electrons. As a result of this interaction, electrons can absorb previously generated plasmons \cite{SR1973} and also excite plasmons that are subsequently outcoupled to cathodoluminescence light emission. The latter is widely used to spectrally and spatially map plasmons and other nanoscale optical modes \cite{paper137,paper149}.

Large electron-photon coupling has been recently accomplished by synchronizing short electron and laser pulses, leading to multiple energy losses and gains, as revealed in the transmitted electron spectra. This technique has been termed photon-induced near-field electron microscopy (PINEM) \cite{BFZ09,YZ12_2,PLQ15,FES15,BZ15}. For the typical beam divergence angles used in electron microscopes ($\sim10$\,mrad), transversal electron motion contributes negligibly to the kinetic energy, so that energy exchanges are mainly due to momentum transfers along the beam direction \cite{paper149}. In fact, an accurate description of the noted spectra only requires dealing with the electron wave function dependence on the path length \cite{paper151}. The evolution along transversal directions has been largely overlooked in this context, although it can be a source of new phenomena, such as angular momentum transfer in the interaction with optical chiral modes \cite{paper243,HPC15}.

In this work, we investigate the phenomenon of electron diffraction by plasmon waves. An electron beam traversing a plasmon standing wave is predicted to undergo strong diffraction in both its elastic and inelastic outgoing components under currently feasible conditions. In contrast to the Kapitza-Dirac effect \cite{KD1933,FAB01,FB02,B07}, mediated by the ponderomotive force of a free-space light wave, electron deflection is produced by the direct action of the electric field associated with the plasmon, without involvement of magnetic fields, and it additionally produces diffracted inelastic electron components. In our predicted electron diffraction effect, the plasmon acts both as a phase grating (by locally modifying the phase of the electron wave function through multiple inelastic exchanges) and as an amplitude grating (by shifting more probability from elastic to inelastic electron beam channels at transversal positions corresponding to maximum plasmon strength). Interestingly, plasmon-induced electron diffraction can be scaled down to ultraconfined modes of pure quasistatic nature, involving small distances, and consequently, relatively large deflection angles.

\section{Theoretical description}

We consider an electron beam of finite lateral extension that passes near an illuminated nanostructure. The electron wave function $\psi(\rb,t)$ evolves according to Schr\"odinger's equation $(H_0+H_1)\psi=\ii\hbar\partial\psi/\partial t$, where $H_0$ is the free-space Hamiltonian and $H_1$ describes the interaction with the optical field. For simplicity, we assume classical monochromatic light of frequency $\omega$. We then have
\begin{align}
H_1=\frac{-e\hbar}{\me\omega}\left(\ee^{-\ii\omega t}\Eb\cdot\nabla-\ee^{\ii\omega t}\Eb^*\cdot\nabla\right),
\label{Hp}
\end{align}
where we use the convention $\Eb(\rb)\ee^{-\ii\omega t}+{\rm c.c.}$ for the electric field amplitude $\Eb$. We envision a metallic nanostructure in which the near field is dominated by induced plasmons, although the present formalism can also be applied to evanescent waves confined to non-plasmonic materials. The electron kinetic energy and momentum are taken to be peaked around $E_0$ and $\hbar k_0\zz$, respectively, where $\zz$ is chosen along the beam direction and $\hbar k_0=\sqrt{2\me E_0}\,\sqrt{1+E_0/2\me c^2}$. We contemplate a small spread in electron kinetic energy ($\ll\hbar\omega$) \cite{KD1}, which is dominated by plane wave components with wave vectors $\kb$ such that $|\kb-k_0\zz|\ll k_0$, each of them satisfying $H_0\ee^{\ii\kb\cdot\rb}=E_\kb\ee^{\ii\kb\cdot\rb}$ in free space. For such narrow $\kb$ distribution, we can approximate $E_\kb\approx E_0+\hbar v(k_z-k_0)$, where $v=(\hbar k_0/\me)/(1+E_0/\me c^2)$ is the peak electron velocity. This relation holds even after interaction with the optical field, under the assumption that $\hbar\omega\ll E_0$ (i.e., neglecting recoil). The unperturbed Hamiltonian can thus be approximated by $H_0\approx E_0-\hbar v\left(\ii\partial/\partial z+k_0\right)$, which suggests that we recast the electron wave function as
\begin{align}
\psi(\rb,t)=\ee^{\ii(k_0z-E_0t/\hbar)}\phi(\rb,t).
\nonumber
\end{align}
Additionally, we replace $\nabla$ by $\ii k_0\zz$ in Eq. (\ref{Hp}) (i.e., we ignore wave function gradients other than the contribution from $\ee^{\ii k_0z}$), so that the Schr\"odinger equation reduces to
\begin{align}
\frac{-ev\gamma}{\hbar\omega}\left(\ee^{-\ii\omega t}\mathcal{E}_z-\ee^{\ii\omega t}\mathcal{E}_z^*\right)\phi=\left(v\frac{\partial}{\partial z}+\frac{\partial}{\partial t}\right)\phi,
\label{Sch2}
\end{align}
where $\gamma=1/\sqrt{1-v^2/c^2}$. Now, in the absence of any interaction ($\mathcal{E}_z=0$), an incident wave function of the form $\phi(\rb,t)=\phi_0(\rb-v\zz t)$ automatically satisfies Eq.\ (\ref{Sch2}). Obviously, this only holds for small distances relative to the interaction region, as diffraction becomes important during free propagation to the far field (see below).

\begin{figure*}
\begin{center}
\includegraphics[width=120mm,angle=0,clip]{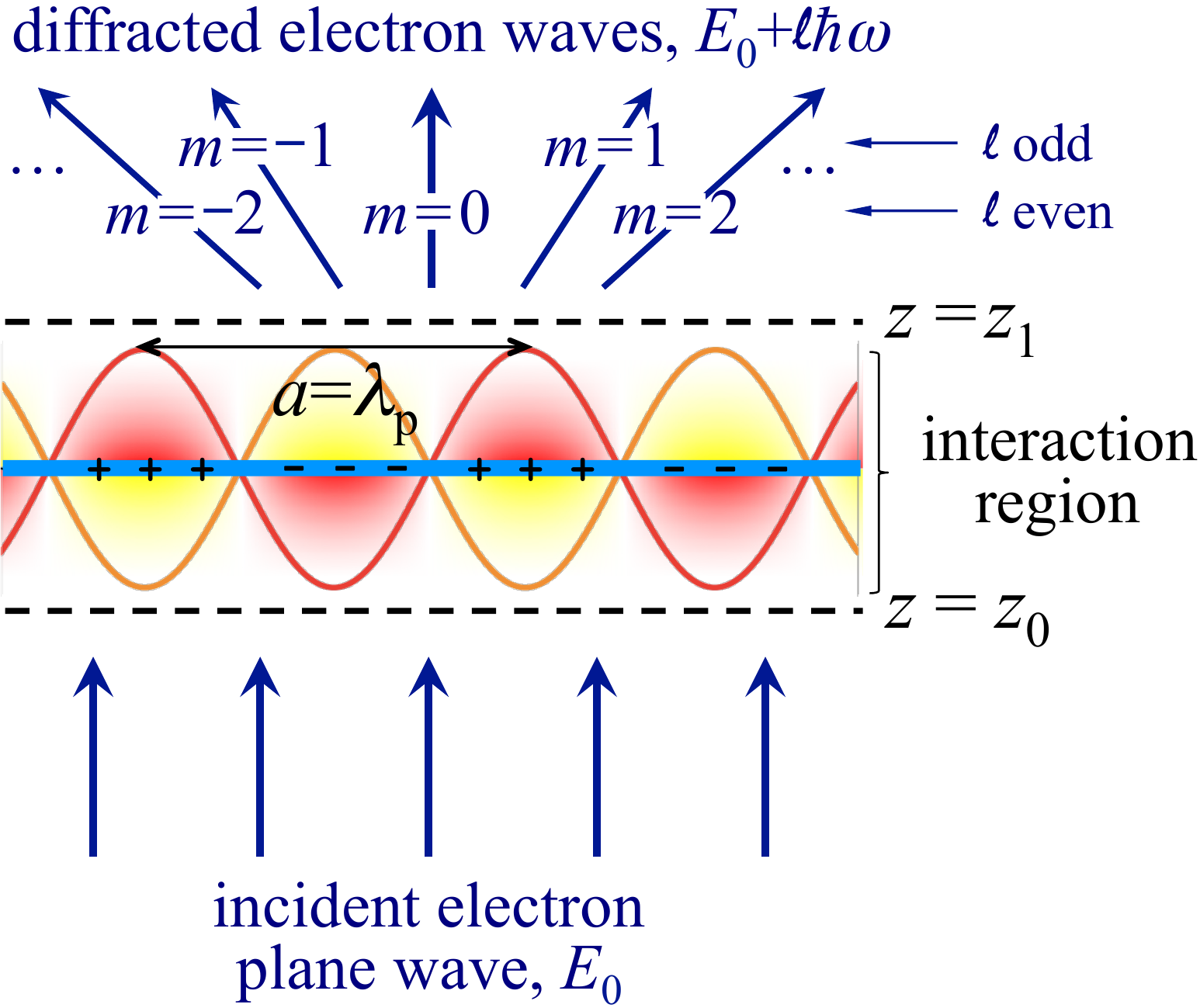}
\caption{Schematic representation of the physical process under consideration. An electron beam (e.g., a plane wave) interacts with a plasmon wave, giving rise to diffracted electron components. For a standing wave formed by two counter-propagating plasmons, the diffracted electrons move along directions determined by the period $a$, which is equal to the plasmon wavelength $\lambdap$. The scattered electron energies differ from the incident energy $E_0$ by a multiple of the plasmon energy $\hbar\omega$.}
\label{Fig1}
\end{center}
\end{figure*}

Photon exchanges between the electron and the optical field modifies the electron energy in steps of $\pm\hbar\omega$. As a result, after several such exchanges, the electron wave function should be distributed among components grouped around a periodically spaced set of energies  $E_0+\lll\hbar\omega$. We actually find that
\begin{align}
\phi(\rb,t)=\phi_0(\rb-v\zz t)\sum_\lll\ee^{\ii\lll\omega(z/v-t)}f_\lll(\rb),
\nonumber
\end{align}
is a solution of Eq.\ (\ref{Sch2}), provided that the space-dependent coefficients $f_\lll(\rb)$ evolve along $z$ according to
\begin{align}
\frac{\partial f_\lll}{\partial z}=\frac{e\gamma}{\hbar\omega}\left(\mathcal{E}_z^*\ee^{\ii\omega z/v}f_{\lll+1}-\mathcal{E}_z\ee^{-\ii\omega z/v}f_{\lll-1}\right).
\label{fp}
\end{align}
Interestingly, Eq.\ (\ref{fp}) does not mix different $\Rb$'s, and therefore, it locally preserves the electron probability
\begin{align}
\sum_\lll|f_\lll(\rb)|^2=1
\label{suml}
\end{align}
as a consequence of the Hermitian character of the corresponding secular matrix.

Before the electron enters the interaction region (see $z=z_0$ plane in Fig.\ \ref{Fig1}), we must have $\phi=\phi_0$, or equivalently, $f_\lll=\delta_{\lll0}$. We then propagate this starting value via Eq.\ (\ref{fp}) up to a plane $z=z_1$ situated right after passing the interaction region (see Fig.\ \ref{Fig1}), where the $\lll^{\rm th}$ energy component of the wave function reduces to $\phi_0(\Rb,z_1-vt)f_\lll(\Rb,z_1)\ee^{\ii\varphi_\lll}$. Here, $\varphi_\lll=k_0z_1-E_0t/\hbar+\lll\omega(z_1/v-t)$ is a global phase and we introduce the notation $\rb=(\Rb,z)$. Beyond that plane, $f_\lll(\rb)$ no longer depends on $z$, so that free-space propagation takes place as we describe below.

Incidentally, following similar methods as in Ref.\ \cite{FES15}, we derive an analytical solution of Eq.\ (\ref{fp}) in the Supplementary Information (SI) \cite{EPAPSKD}, where we find $f_\lll(\beta)=\ee^{\ii\lll\arg\{-\beta\}}\,J_{\lll}(2|\beta|)$ in terms of Bessel functions and the integrated amplitude
\[\beta(z)=\frac{e\gamma}{\hbar\omega}\int_{-\infty}^z dz'\, \mathcal{E}_z(z')\,\ee^{-\ii\omega z'/v}.\]
Equation\ (\ref{suml}) is then trivially satisfied from the property $\sum_\lll J_\lll^2(x)=1$ for any real argument $x$. Notice these expressions are valid for any point $z$ along the evolution of the wave function.

\section{Electron cross-section of a localized plasmon}

Consider an incident monochromatic electron plane wave ($\phi_0=1/\sqrt{V}$, where $V$ is the normalization volume) interacting with a localized plasmon. Scalar diffraction theory \cite{J99} allows us to write, for $z>z_1$,
\begin{align}
\psi(\rb,t)=\frac{1}{4\pi^2\sqrt{V}}\sum_\lll\ee^{\ii\varphi_\lll}
\int d^2\kperb \ee^{\ii\kb_\lll\cdot\rb} f_{\lll,\kperb},
\label{psidiff}
\end{align}
where $f_{\lll,\kperb}=\int d^2\Rb\,\ee^{-\ii\kb_\perp\cdot\Rb} f_\lll(\Rb,z_1)$, $\kb_\lll=(\kperb,k_{z\lll})$, and $k_{z\lll}\approx\sqrt{(k_0+\lll\omega/v)^2-k_\perp^2+\ii0^+}$ (with ${\rm Im}\{k_{z\lll}\}>0$). We are interested in the far-field limit ($k_0r\gg1$), where the wave function reduces to $\psi(\rb,t)\approx(-\ii k_z\ee^{\ii k_0r}/2\pi r\sqrt{V})
\sum_\lll\ee^{\ii\varphi_\lll}f_{\lll,\kperb}$, with $\kb\approx k_0\rr$. From this expression, we calculate the electron current $(\hbar/\me){\rm Im}\{\psi^*\nabla\psi\}$ by approximating $\nabla\approx\ii \kb$. We then divide the result by the incident current $v\gamma/\sqrt{V}$ to obtain the scattering cross-section $\sigma=\sum_\lll\int d\Omega \,\sigma_\lll(\rr)$, whose partial contributions
\begin{align}
\sigma_\lll(\rr)=\frac{k_z^2}{4\pi^2}\left|\int d^2\Rb\,\ee^{-\ii\kb\cdot\Rb}f_\lll(\Rb,z_1)\right|^2
\nonumber
\end{align}
are separated in components $\lll$ and outgoing directions $\rr$.

\section{Diffraction by a periodic plasmon wave}

For simplicity, we focus on an electron plane wave normally traversing a plasmon standing wave that is invariant along $y$ and periodic along $x$ with period $a$ (see Fig.\ \ref{Fig1}). The transmitted electron consists of a discrete set of beams labeled by both the net number of exchanged plasmons $\lll$ and the diffraction order $m$. In the far field, using Eq.\ (\ref{psidiff}) and assuming small scattering angles $2\pi|m|/k_0a\ll1$, we find the corresponding currents
\begin{align}
I_{\lll m}\approx I_{\rm inc}\left|\frac{1}{a}\int_0^a dx\,\ee^{-2\pi\ii mx/a}f_\lll(x,0,z_1)\right|^2,
\label{Ilm}
\end{align}
where $I_{\rm inc}$ is the incident current. It is reassuring to observe that, as a consequence of Eq.\ (\ref{suml}), the total current $\sum_{\lll m}I_{\lll m}=I_{\rm inc}$ is preserved.

In a possible realization of this idea, we consider a self-standing thin conductive film characterized by its 2D optical conductivity $\sigma$. Neglecting optical losses (i.e., ${\rm Re}\{\sigma\}\ll{\rm Im}\{\sigma\}$), the film supports plasmons of wavelength $\lambdap=4\pi^2{\rm Im}\{\sigma\}/\omega$, subject to the condition ${\rm Im}\{\sigma\}>0$. In particular, plasmons in thin noble metals \cite{paper254} and highly doped graphene \cite{paper235} exhibit small wavelengths $\lambdap\ll c/\omega$, which allow us to safely work within the quasistatic limit.

Two counter-propagating plasmon waves set up a periodic electric field of period $a=\lambdap$ (see Fig.\ \ref{Fig1}) whose $z$ component reduces to
\begin{align}
\mathcal{E}_z =\mathcal{E}_0\,{\rm sign}(z)\,\ee^{-2\pi|z|/\lambdap}\cos(2\pi x/\lambdap).
\label{Ez}
\end{align}
Upon examination of Eq.\ (\ref{fp}) [see SI \cite{EPAPSKD}, where we derive analytical solutions], we conclude that the outgoing diffracted beam intensities produced by this field only depend on two parameters: the normalized plasmon amplitude
\begin{align}
g=e\mathcal{E}_0\lambdap\gamma/2\pi\hbar\omega\approx e\mathcal{E}_0\lambdap/2\pi\hbar\omega,
\label{eqg}
\end{align}
where the rightmost approximation stands for nonrelativistic electrons, and
\begin{align}
\Omega=\omega\lambdap/2\pi v,
\label{eqOmega}
\end{align}
which represents the number of optical cycles taken by the electron to move along a distance $\lambdap$.

Remarkably, as we prove in the SI (see Eq.\ (S8) in Ref.\ \cite{EPAPSKD}), $I_{\lll m}$ depends on $g$ and $\Omega$ only through the parameter $|\eta|=2|g\Omega|/(1+\Omega^2)$, it vanishes if $\lll+m$ is odd, it exhibits the symmetry $I_{\lll m}=I_{m\lll}$, and it is independent of the signs of $\lll$ and $m$. Further inspection reveals that $I_{lm}$ scales as $|\eta|^{2\max\{|\lll|,|m|\}}$ for small $\eta$.

\begin{figure*}
\begin{center}
\includegraphics[width=100mm,angle=0,clip]{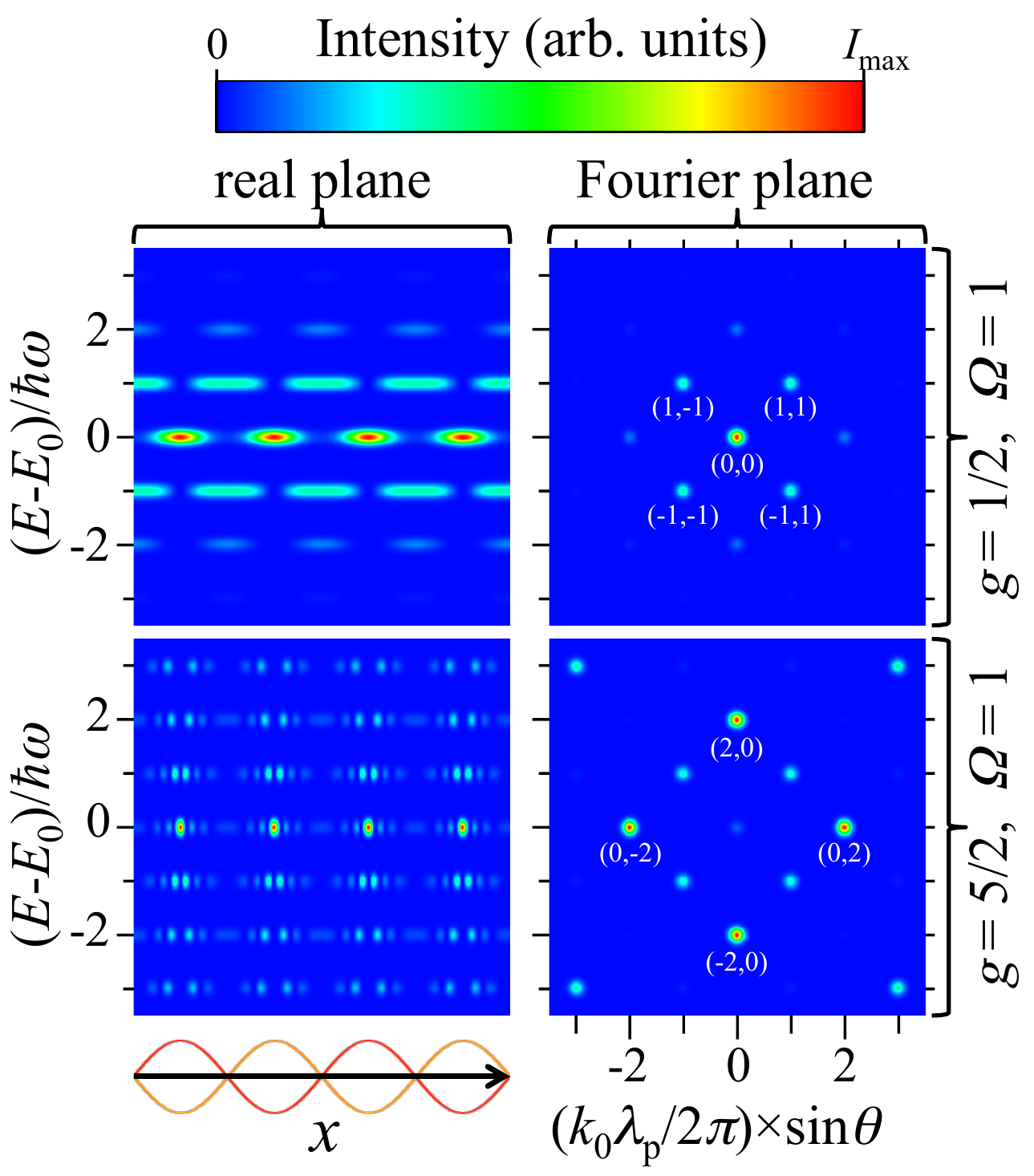}
\caption{Electron intensity upon interaction of an electron plane wave with a standing plasmon wave in the real (left) and Fourier (right) planes for a normalized plasmon frequency $\Omega=1$ and two different plasmon amplitudes, $g=1/2$ (top) and $g=5/2$ (bottom). An energy broadening $\sim\hbar\omega/5$ has been introduced for the sake of clarity in all plots. Additionally, the Fourier plots have been broaded with a transversal momentum uncertainty $\sim2\pi/5\lambdap$, and some of the main features labeled with the corresponding $(\lll,m)$ values. The symmetries $I_{\lll m}=I_{m\lll}=I_{-\lll m}=I_{\lll,-m}$ are apparent.}
\label{Fig2}
\end{center}
\end{figure*}

Right after passing the plasmon region (at $z=z_1$), the $x$-dependent intensity of the different $\lll$ electron components is directly given by $|f_\lll|^2$, as illustrated in Fig.\ \ref{Fig2} (left) for $\Omega=1$ and two different values of the plasmon amplitude $g$. These plots directly corresponds to recently measured energy- and space-resolved plasmon standing waves using PINEM \cite{PLQ15}. They also reveal a complex dependence of the intensity on local plasmon field amplitude, which gives rise to migration of the wave function to adjacent $\lll$ channels during the interaction (see Eq.\ (\ref{fp}) and Refs.\ \cite{paper151,FES15}), an effect that we illustrate in Fig.\ S3 of the SI \cite{EPAPSKD}. The evolution toward the detector in the far field (Fourier plane) yields electron intensities that depend on the deflection angle $\theta$ as shown in the right plots of Fig.\ \ref{Fig2}, where each spot corresponds to a given choice of $\lll$ and $m$, centered around angles determined by $k_0\sin\theta\approx2\pi m/\lambdap$. The real (Fourier) plane images shown in Fig.\ \ref{Fig2} correspond to what one would observe with focused (plane-wave) electron beams by rastering the spatial (angular) coordinate $x$ ($\theta$).

More colorful results are expected in the opposite regime of large optical intensities. In order to estimate the maximum range of $g$ and $\Omega$ under attainable experimental conditions, we note that the applied light intensity can reach $\sim\,$GW/cm$^2$ \cite{BFZ09,YZ12_2,PLQ15,FES15}, while coupling to plasmons can produce additional enhancement in the resulting field amplitude. Considering a plasmon of wavelength $\lambdap\sim100\,$nm \cite{KD3} and energy $\hbar\omega\sim1\,$eV, we find values of $g>20$. With these parameters, we also find $\Omega\sim0.1-4$ for electrons in the $0.1-200\,$keV energy range, which covers a wide selection of available electron microscope regimes.

\begin{figure*}
\begin{center}
\includegraphics[width=100mm,angle=0,clip]{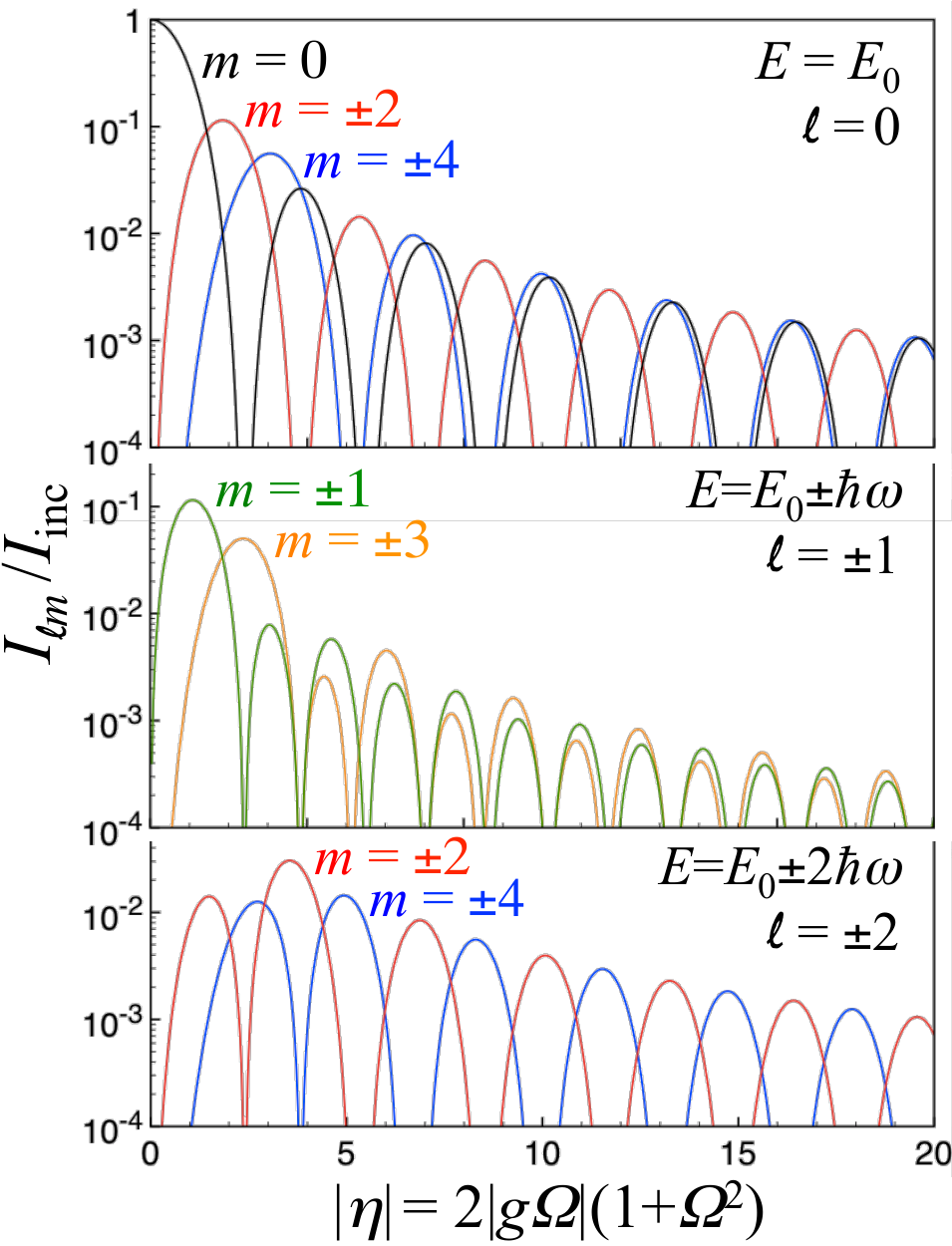}
\caption{Electron current $I_{\lll m}$ [Eq.\ (\ref{Ilm})] for different diffraction orders $m$ and outgoing energies $E_0+\lll\hbar\omega$ under the conditions of Fig.\ \ref{Fig1} as a function of $|\eta|=2|g\Omega|/(1+\Omega^2)$.}
\label{Fig3}
\end{center}
\end{figure*}

The $\eta$ dependence of $I_{\lll m}$ is shown in Fig.\ \ref{Fig3} for a few low-order beams, as directly obtained from our analytical solution of Eqs.\ (\ref{fp}) and (\ref{Ilm}) (see Eq.\ (S8) in Ref. \cite{EPAPSKD}). These results are supplemented with more detailed plots of both $f_\lll$ and $I_{\lll m}$ in Figs.\ S1 and S2 of the SI \cite{EPAPSKD}. We observe that the incident beam ($\lll=m=0$) is initially depleted as the plasmon strength increases. Full depletion, which takes place at $\eta\sim2.4$, indicates an optimum electron-plasmon coupling, accompanied by relatively large intensities of the rest of the beams ($\lll\neq0$ or $m\neq0$), and eventually followed by successive partial revivals for increasing $\eta$. For small $g$, optimum coupling occurs at $\Omega=1$ (maximum of $\eta$), as qualitatively expected by the following argument: the plasmon field changes sign across the film [see Eq.\ (\ref{Ez})]; however, its temporal evolution [$\ee^{\pm\ii\omega z/v}$ factors of Eq.\ (\ref{fp})] can be timed in such a way that they change sign (for $\omega z/v=\pi$) over a distance comparable to the spill out of the plasmon away from the film ($z\sim\lambdap/2\pi$); when this happens, the electric force experienced by the electron adds up on both sides of the film, giving rise to maximum interaction. Interestingly, $I_{02}=I_{20}$ and $I_{11}$ can take values exceeding 10\%.

In practice, a normally incident laser pulse (duration $\Delta_{\rm p}$)  could launch the plasmons by interacting with two parallel slits that are separated by a distance $d$ across a flat film region. An in-coincidence electron pulse (duration $\Delta_{\rm e}$) then feels the resulting plasmon standing wave \cite{KD2}. A realistic choice of parameters similar to those in recent experiments \cite{BFZ09,YZ12_2,PLQ15,FES15,KD2} (e.g., $\Delta_{\rm p}\sim1000\times2\pi/\omega$, $\Delta_{\rm e}\sim100\times2\pi/\omega$, and $d\sim10\lambdap$) satisfies the conditions that are necessary (1) for the plasmon standing wave to be well developed ($\omega\Delta_{\rm p}/2\pi\gg1$), (2) for the electron to see a nearly stationary plasmon regime ($\Delta_{\rm p}\gg\Delta_{\rm e}\gg2\pi/\omega$), and (3) for the plasmon {\it grating} to exhibit a sufficient number of periods ($d\gg\lambdap$) as to produce a few well-defined low-$m$ diffraction orders. Additionally, as the electron wavelength is small compared with $\lambdap$, an incident electron beam spanning a lateral size $d$ should present a negligible angular divergence \cite{KD3}. Finally, the plasmon propagation distance has to be large compared with $d$, a condition that is satisfied in high-quality graphene \cite{WLG15}. Incidentally, strong collisions with atoms in the plasmon-supporting film and inelastic scattering by phonons and other excitations could result in an electron background distribution, similar to what happens in EELS and electron holography experiments.

\section{Concluding remarks}

We have focused on a specific realization of electron diffraction by plasmon waves, but there are many other possible geometries. For example, one could exploit high-order plasmon standing waves of long silver nanowires such as those recently synthesized and imaged by space-resolved EELS \cite{paper258}. Plasmon snapshots similar to those of Fig.\ 2 (left) have been already recorded in this geometry under plane wave illumination \cite{PLQ15}, whereas here we propose a Fourier plane analysis. Such nanowire geometry is compatible with an aloof beam configuration, in which the electron is not directly trespassing any material boundary, thus avoiding undesired strong collisions with target atoms, or coupling to inelastic modes (e.g., phonons). Alternatively, electrons could be electrostatically deflected from a biased surface \cite{LBN1977}, on which a standing plasmon wave would provide the means to produce efficient electron diffraction, thus addressing a pending challenge in the coherent manipulation of low-energy free electron waves \cite{B07}. Active control of electron diffraction also appears to be possible by playing with the frequency, intensity, and symmetry of the light used to excite the plasmons, resulting in engineered plasmon waves that could for instance produce diffracted vortex electron beams \cite{HSB15}. The ultrafast dynamics of plasmons adds the possibility of shaping electron diffraction with high temporal resolution. We also note that other polaritonic excitations (e.g., optical phonons in 2D crystals) could be employed instead of plasmons. These ideas constitute the basis for the development of low-energy electron optics setups in which the electrons interact with evanescent light fields without the damaging effect of close encounters with bulky materials.

\acknowledgments

This work has been supported in part by the Spanish MINECO (MAT2014-59096-P and SEV-2015-0522) and the Suisse National Science Foundation (NCCR grant MUST).

\appendix

We derive an analytical solution for the wave function coefficients $f_\lll$ and the beam currents $I_{\lll m}$ from which several interesting symmetry properties are obtained. Furthermore, we provide graphical results that supplement those of the main paper.

\section{Analytical solution for $f_\lll$}

We seek to find an analytical solution for the electron wave function coefficients $f_\lll$, whose evolution is described by Eq.\ (3) of the main paper. It is convenient to define a dimensionless path length coordinate $\theta=z/D$, where $D$ is a characteristic length of the system (e.g., $D=\lambdap/2\pi$ for a plasmon grating). We also separate the electric field as
\begin{align}
\mathcal{E}_z=\mathcal{E}_0\,A(\theta),
\label{Ez}
\end{align}
where $A(\theta)$ is a dimensionless function. This allows us to recast Eq.\ (3) as
\begin{align}
\frac{\partial f_\lll(\theta)}{\partial\theta}=g\left[A^*(\theta)\ee^{\ii\Omega\theta}f_{\lll+1}(\theta)-A(\theta)\ee^{-\ii\Omega\theta}f_{\lll-1}(\theta)\right],
\label{fp}
\end{align}
where
\begin{align}
g=e\mathcal{E}_0D\gamma/\hbar\omega
\nonumber
\end{align}
and
\begin{align}
\Omega=\omega D/v
\nonumber
\end{align}
are defined by analogy to Eqs.\ (8) and (9). This system of coupled-channel equations has been solved in exact form by using a second-quantization formalism \cite{FES15}. Here, we provide a similar derivation that is suitable to later study electron diffraction, and that, in contrast to Ref.\  \cite{FES15}, yields the electron amplitude, the squared modulus of which is the electron intensity.

We can now write Eq.\ (\ref{fp}) in matrix form as
\begin{align}
\frac{\partial f(\theta)}{\partial\theta}=g\left[
A^*(\theta)\ee^{\ii\Omega\theta} U-A(\theta)\ee^{-\ii\Omega\theta}U^{-1}\right]\cdot f(\theta),
\nonumber
\end{align}
where $f$ is the vector of components $f_\lll$, whereas $U$ is a matrix of coefficients $U_{\lll\lll'}=\delta_{\lll,\lll'-1}$, and obviously, $(U^{-1})_{\lll\lll'}=\delta_{\lll,\lll'+1}$. Noticing that all matrices formed as linear combinations of $U$ and $U^{-1}$ commute among themselves, we integrate the above differential equation to find
\begin{align}
f(\theta)=\exp\left\{g\int_{\theta_0}^{\theta}d\theta'\left[A^*(\theta')\ee^{\ii\Omega\theta'} U-A(\theta')\ee^{-\ii\Omega\theta'}U^{-1}\right]\right\}\cdot f(\theta_0).
\nonumber
\end{align}
In particular, the outgoing coefficients after interaction with the plasmon field reduce to
\begin{align}
f(\theta)=\ee^{\beta^*(\theta)U-\beta(\theta) U^{-1}}\cdot f(-\infty),
\nonumber
\end{align}
where
\begin{align}
\beta(\theta)=g\int_{-\infty}^\theta d\theta'\,A(\theta')\,\ee^{-\ii\Omega\theta'}.
\label{beta}
\end{align}
For an incident plane wave, we use $f_\lll(-\infty)=\delta_{\lll,0}$ to write the analytical solution
\begin{align}
f_\lll(\infty)&=\ee^{\beta^*U-\beta U^{-1}}\big|_{\lll0} \nonumber\\
&=\sum_{n=0}^{\infty}\frac{1}{n!}\left[\left(\beta^*U-\beta U^{-1}\right)^n\right]_{\lll0} \nonumber\\
&=\sum_{n=0}^{\infty}\sum_{j=0}^{n}\frac{(\beta^*)^{n-j}(-\beta)^j}{(n-j)!\,j!}\left(U^{n-2j}\right)_{\lll0} \nonumber\\
&=\sum_{n=0}^{\infty}\sum_{j=0}^{n}\frac{(\beta^*)^{n-j}(-\beta)^j}{(n-j)!\,j!}\delta_{\lll,2j-n} \nonumber\\
&=\sum_{j=0}^{\infty}\sum_{n=j}^{\infty}\frac{(\beta^*)^{n-j}(-\beta)^j}{(n-j)!\,j!}\delta_{\lll,2j-n} \nonumber\\
&=(\beta^*)^{-\lll}\sum_{j=\max\{0,\lll\}}^{\infty}\frac{(-1)^j|\beta|^{2j}}{(j-\lll)!\,j!} \nonumber\\
&=\sum_{j=0}^{\infty}\frac{(-1)^j|\beta|^{2j}}{(|\lll|+j)!\,j!}\times
   \begin{cases}
   (-\beta)^{\lll},\;\;\;\;\lll\ge0, \\
   (\beta^*)^{-\lll},\;\;\;\lll<0,
   \end{cases} \nonumber\\
&= \ee^{\ii\lll\arg\{\beta\}}\,J_{|\lll|}(2|\beta|)\times
   \begin{cases}
   (-1)^\lll,\;\;\;\lll\ge0, \\
  1,\;\;\;\;\;\;\;\;\;\;\lll<0,
  \end{cases} \nonumber\\
&=\ee^{\ii\lll\arg\{-\beta\}}\,J_{\lll}(2|\beta|)
\label{flsol}
\end{align}
where we have inserted the binomial expansion of $\left(\beta^*U-\beta U^{-1}\right)^n$ and identified the penultimate line with the Taylor expansion of the Bessel function $J_\lll(x)=(x/2)^\lll\sum_j(-x^2/2)^j/j!(j+\lll)!$ \cite{AS1972}. Remarkably, $f_\lll$ only depends on the integrated field amplitude $\beta\propto g$ [see Eq.\ (\ref{beta})]. Additionally, we see that the intensity of the $\lll$ outgoing electron component is $|f_\lll|^2=|\beta|^{2|\lll|}/(|\lll|!)^2+O(|\beta|^{2|\lll|+2})\propto g^{2|\lll|}$.

\begin{figure*}
\begin{center}
\includegraphics[width=140mm,angle=0,clip]{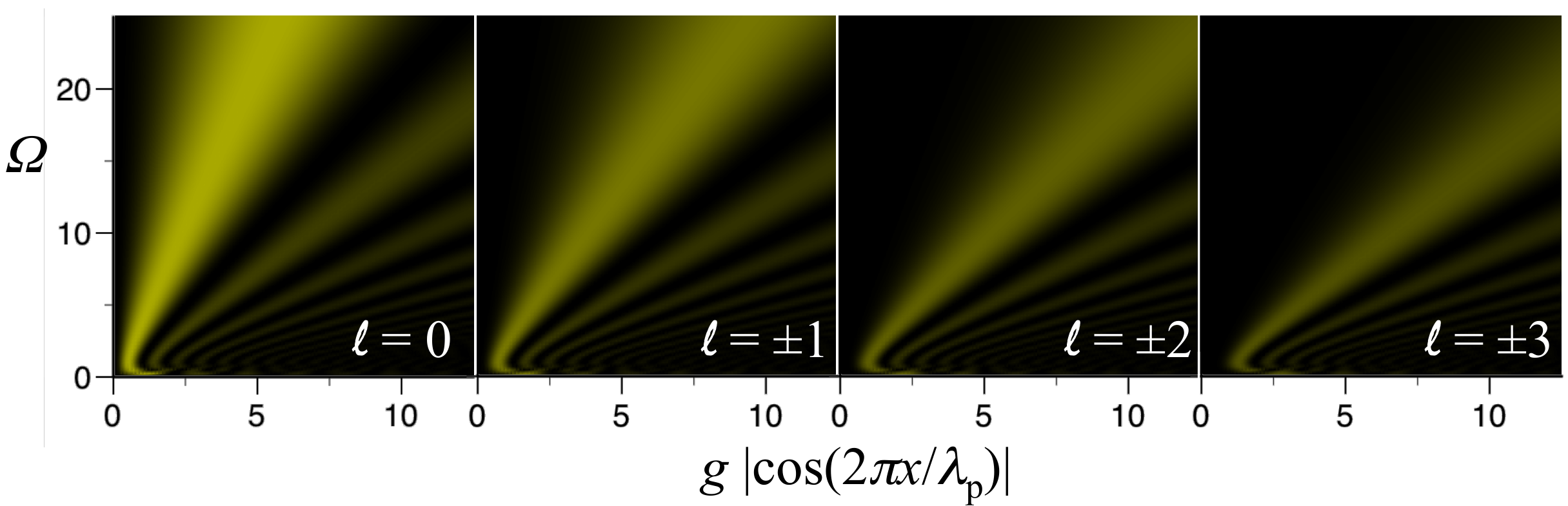}
\caption{Electron wave function coefficients $|f_\lll|^2$ right after electron-plasmon interaction under the same conditions as in Fig.\ 1 of the main paper. We show the dependence of these coefficients on the normalized plasmon frequency $\Omega$ and the $x$-dependent plasmon amplitude $g\cos(2\pi x/\lambdap)$ for $|\lll|\le3$ [see Eqs.\ (8) and (9)].}
\label{FigSI1}
\end{center}
\end{figure*}

\begin{figure*}
\begin{center}
\includegraphics[width=100mm,angle=0,clip]{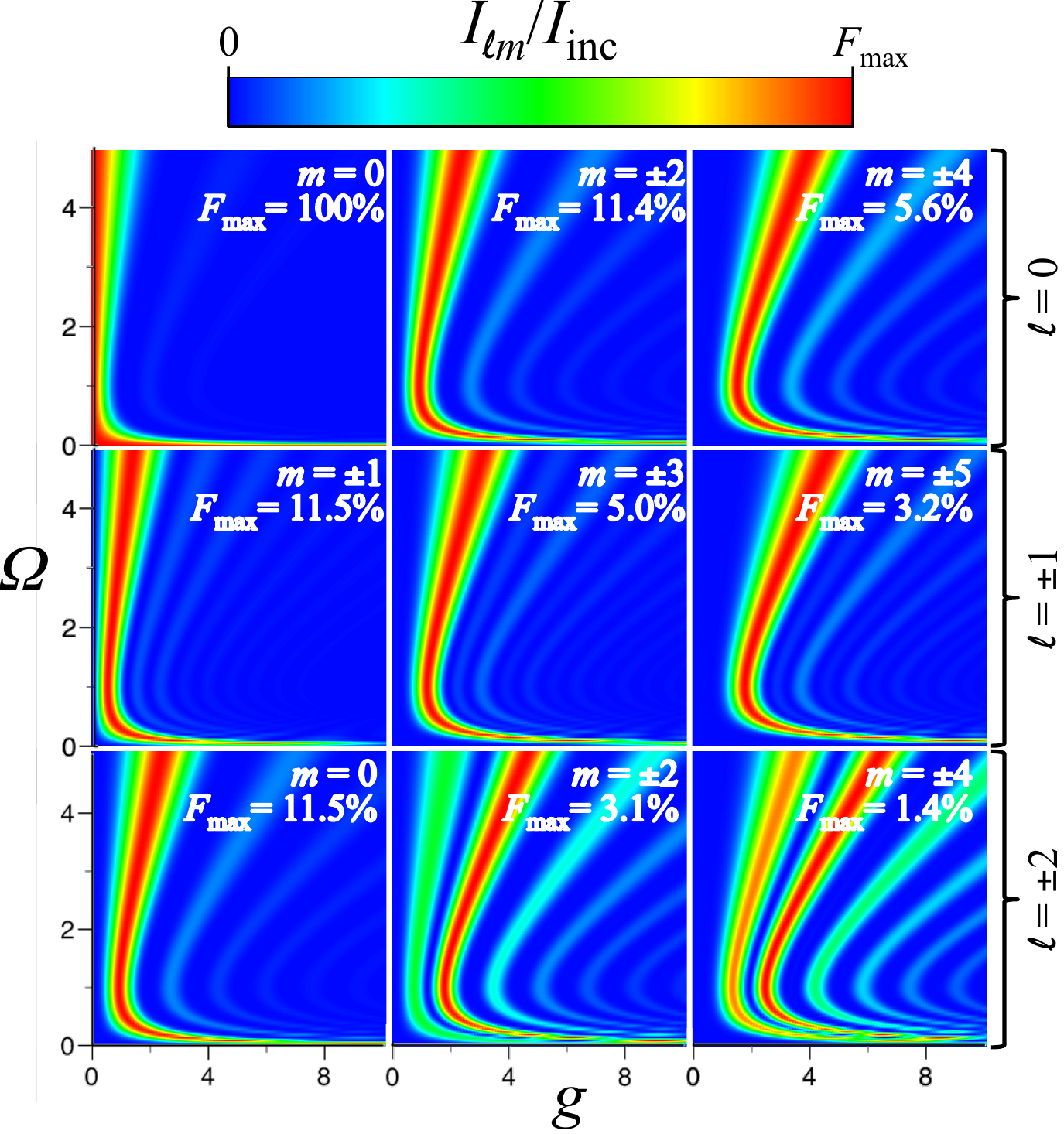}
\caption{Electron current $I_{\lll m}$ for different diffraction orders $m$ and outgoing energies $E_0+\lll\hbar\omega$ under the conditions of Fig.\ 1 of the main paper, as a function of the normalized plasmon amplitude $g$ and frequency $\Omega$ [see Eqs.\ (8) and (9)]. The maximum percentage of beam fraction is indicated by labels in each plot.}
\label{FigSI2}
\end{center}
\end{figure*}

\section{$I_{\lll m}$ for electron diffraction by a plasmon grating}

For the plasmon grating considered in the main paper, we set $D=\lambdap/2\pi$ and $A(\theta)={\rm sign}(\theta)\,\ee^{-|\theta|}\cos(x/D)$ [cf. Eqs. (\ref{Ez}) and (7)]. This yields the integrated amplitude
\begin{align}
\beta=-\ii\eta\cos(x/D),
\label{betabis}
\end{align}
where $\eta=2g\Omega/(1+\Omega^2)$. We now calculate the diffracted beam intensities from Eq.\ (6) of the main paper as $I_{\lll m}\approx I_{\rm inc}|a_{\lll m}|^2$, where
\begin{align}
a_{\lll m}=\frac{1}{2\pi D}\int_0^{2\pi D} dx\,\ee^{-\ii mx/D}f_\lll
\label{alm}
\end{align}
are the corresponding beam amplitudes. This integral requires the evaluation of the Fourier transform of the powers of $\cos(x/D)$,
\begin{align}
&\frac{1}{2\pi D}\int_0^{2\pi D} dx\,\ee^{-\ii mx/D}\cos^n(x/D) \nonumber\\
=&\frac{1}{2^n2\pi D}\int_0^{2\pi D} dx\,\ee^{-\ii mx/D}\left(\ee^{\ii x/D}+\ee^{-\ii x/D}\right)^n \nonumber\\
=&\frac{1}{2^n}\sum_{j'=0}^{n}\frac{n!}{(n-j')!\,j'!}\frac{1}{2\pi D}\int_0^{2\pi D} dx\,\ee^{\ii(n-2j'-m)x/D} \nonumber\\
=&\frac{1}{2^n}\sum_{j'=0}^{n}\frac{n!}{(n-j')!\,j'!}\delta_{n-2j',m},
\label{cosn}
\end{align}
where we have used again a binomial expansion. Combining the second last line of Eq.\ (\ref{flsol}) with Eqs.\ (\ref{betabis})$-$(\ref{cosn}), and taken $\eta$ to be real and positive without loss of generality, we find
\begin{align}
a_{\lll m}=\ii^{|\lll|}\sum_{j=0}^{\infty}\sum_{j'=0}^{|\lll|+2j}\frac{(-1)^j(\eta/2)^{|\lll|+2j}(|\lll|+2j)!}{(|\lll|+j)!\,j!(m+j')!\,j'!}\delta_{|\lll|+2j,m+2j'}.
\nonumber
\end{align}
The factor $\delta_{\lll+2j-2j',m}$ in this expression makes it clear that $a_{\lll m}=0$ if $\lll+m$ is an odd number. Finally, when $\lll+m$ is even, we have
\begin{align}
a_{\lll m}=\sum_{n=0}^{\infty}\frac{(\ii\eta/2)^{2n+N}(2n+N)!}{[n+(N+\lll)/2]![n+(N-\lll)/2]![n+(N+m)/2]![n+(N-m)/2]!},
\label{almfinal}
\end{align}
where $N=\max\{|\lll|,|m|\}$. Incidentally, we have verified that Eq.\ (\ref{almfinal}) produces results that cannot be distinguished in the presented figures from those obtained by direct numerical integration of Eqs.\ (3) and (6) of the main paper.

Interestingly, Eq.\ (\ref{almfinal}) has the symmetry $a_{\lll m}=a_{m\lll}$. Additionally, we remark that the beam currents depend on $g$ and $\Omega$ only through $|\eta|=2|g\Omega|/(1+\Omega^2)$. We present in Fig.\ 3 of the main paper an overview of this dependence for the lowest-order beams and supplement those results with plots presented in next section.

\begin{figure*}
\begin{center}
\includegraphics[width=70mm,angle=0,clip]{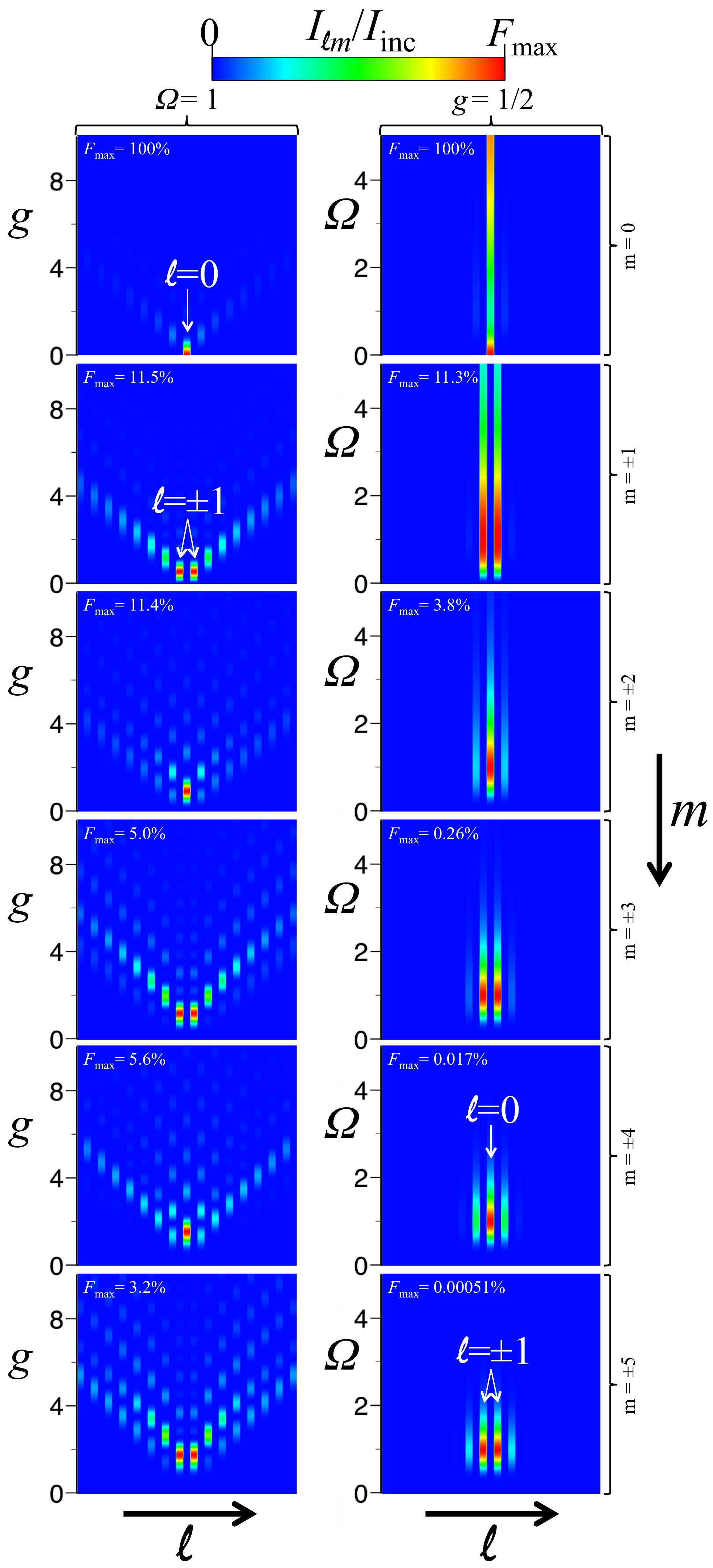}
\caption{Evolution of beam intensity fractions under the same conditions as in Fig.\ 1 of the main paper for different diffracted beams as a function of the net number of exchanged plasmons $\lll$ (horizontal axis, outgoing electron energies equal to $E_0+\lll\hbar\omega$). Each plot corresponds to a given $m$ running from 0 to $\pm5$ (top to bottom). The vertical axes show the dependence on either $g$ (left plots for fixed $\Omega=1$) or $\Omega$ (right plots for fixed $g=1/2$). We note that the intensity is zero for beams in which $\lll+m$ is an even integer.}
\label{FigSI3*}
\end{center}
\end{figure*}

\section{Additional numerical results}

We plot the quantities $|f_\lll|^2$ and $I_{\lll m}$ in Figs.\ \ref{FigSI1} and \ref{FigSI2} as a function of $g$ and $\Omega$ for a few low values of $\lll$ and $m$. Additionally, we show $I_{\lll m}$ in Fig.\ \ref{FigSI3} as a function of the net number of exchanged plasmons $\lll$ (horizontal axes) and either $g$ (left) or $\Omega$ (right).


\begin{thebibliography}{35}
\expandafter\ifx\csname natexlab\endcsname\relax\def\natexlab#1{#1}\fi
\expandafter\ifx\csname bibnamefont\endcsname\relax
  \def\bibnamefont#1{#1}\fi
\expandafter\ifx\csname bibfnamefont\endcsname\relax
  \def\bibfnamefont#1{#1}\fi
\expandafter\ifx\csname citenamefont\endcsname\relax
  \def\citenamefont#1{#1}\fi
\expandafter\ifx\csname url\endcsname\relax
  \def\url#1{\texttt{#1}}\fi
\expandafter\ifx\csname urlprefix\endcsname\relax\def\urlprefix{URL }\fi
\providecommand{\bibinfo}[2]{#2}
\providecommand{\eprint}[2][]{\url{#2}}

\bibitem[{\citenamefont{Kapitza and Dirac}(1933)}]{KD1933}
\bibinfo{author}{\bibfnamefont{P.~L.} \bibnamefont{Kapitza}} \bibnamefont{and}
  \bibinfo{author}{\bibfnamefont{P.~A.~M.} \bibnamefont{Dirac}},
  \bibinfo{journal}{Proc.\ Cambridge\ Philos.\ Soc.}
  \textbf{\bibinfo{volume}{29}}, \bibinfo{pages}{297} (\bibinfo{year}{1933}).

\bibitem[{\citenamefont{Freimund et~al.}(2001)\citenamefont{Freimund,
  Aflatooni, and Batelaan}}]{FAB01}
\bibinfo{author}{\bibfnamefont{D.~L.} \bibnamefont{Freimund}},
  \bibinfo{author}{\bibfnamefont{K.}~\bibnamefont{Aflatooni}},
  \bibnamefont{and} \bibinfo{author}{\bibfnamefont{H.}~\bibnamefont{Batelaan}},
  \bibinfo{journal}{Nature} \textbf{\bibinfo{volume}{413}},
  \bibinfo{pages}{142} (\bibinfo{year}{2001}).

\bibitem[{\citenamefont{Freimund and Batelaan}(2002)}]{FB02}
\bibinfo{author}{\bibfnamefont{D.~L.} \bibnamefont{Freimund}} \bibnamefont{and}
  \bibinfo{author}{\bibfnamefont{H.}~\bibnamefont{Batelaan}},
  \bibinfo{journal}{Phys.\ Rev.\ Lett.} \textbf{\bibinfo{volume}{89}},
  \bibinfo{pages}{283602} (\bibinfo{year}{2002}).

\bibitem[{\citenamefont{Batelaan}(2007)}]{B07}
\bibinfo{author}{\bibfnamefont{H.}~\bibnamefont{Batelaan}},
  \bibinfo{journal}{Rev.\ Mod.\ Phys.} \textbf{\bibinfo{volume}{79}},
  \bibinfo{pages}{929} (\bibinfo{year}{2007}).

\bibitem[{C19()}]{C1934}
\bibinfo{note}{P.~A. Cherenkov, Dokl. Akad. Nauk SSSR {\bf 2}, 451 (1934).}

\bibitem[{FT1()}]{FT1937}
\bibinfo{note}{I.~M. Frank and I. Tamm, Dokl. Akad. Nauk SSSR {\bf 14}, 109
  (1937).}

\bibitem[{\citenamefont{Ginzburg}(1996)}]{G96}
\bibinfo{author}{\bibfnamefont{V.~L.} \bibnamefont{Ginzburg}},
  \bibinfo{journal}{Phys.\ Usp.} \textbf{\bibinfo{volume}{39}},
  \bibinfo{pages}{973} (\bibinfo{year}{1996}).

\bibitem[{\citenamefont{Edighoffer et~al.}(1981)\citenamefont{Edighoffer,
  Kimura, Pantell, Piestrup, and Wang}}]{EKP1981}
\bibinfo{author}{\bibfnamefont{J.~A.} \bibnamefont{Edighoffer}},
  \bibinfo{author}{\bibfnamefont{W.~D.} \bibnamefont{Kimura}},
  \bibinfo{author}{\bibfnamefont{R.~H.} \bibnamefont{Pantell}},
  \bibinfo{author}{\bibfnamefont{M.~A.} \bibnamefont{Piestrup}},
  \bibnamefont{and} \bibinfo{author}{\bibfnamefont{D.~Y.} \bibnamefont{Wang}},
  \bibinfo{journal}{Phys.\ Rev.\ A} \textbf{\bibinfo{volume}{23}},
  \bibinfo{pages}{1848} (\bibinfo{year}{1981}).

\bibitem[{\citenamefont{Kimura et~al.}(1995)\citenamefont{Kimura, Kim, Romea,
  Steinhauer, Pogorelsky, Kusche, Fernow, Wang, and Liu}}]{KKR95}
\bibinfo{author}{\bibfnamefont{W.~D.} \bibnamefont{Kimura}},
  \bibinfo{author}{\bibfnamefont{G.~H.} \bibnamefont{Kim}},
  \bibinfo{author}{\bibfnamefont{R.~D.} \bibnamefont{Romea}},
  \bibinfo{author}{\bibfnamefont{L.~C.} \bibnamefont{Steinhauer}},
  \bibinfo{author}{\bibfnamefont{I.~V.} \bibnamefont{Pogorelsky}},
  \bibinfo{author}{\bibfnamefont{K.~P.} \bibnamefont{Kusche}},
  \bibinfo{author}{\bibfnamefont{R.~C.} \bibnamefont{Fernow}},
  \bibinfo{author}{\bibfnamefont{X.}~\bibnamefont{Wang}}, \bibnamefont{and}
  \bibinfo{author}{\bibfnamefont{Y.}~\bibnamefont{Liu}},
  \bibinfo{journal}{Phys.\ Rev.\ Lett.} \textbf{\bibinfo{volume}{74}},
  \bibinfo{pages}{546} (\bibinfo{year}{1995}).

\bibitem[{\citenamefont{{Garc\'{\i}a de Abajo}
  et~al.}(2003)\citenamefont{{Garc\'{\i}a de Abajo}, {A. G.
  Pattantyus-Abraham}, Zabala, Rivacoba, Wolf, and Echenique}}]{paper048}
\bibinfo{author}{\bibfnamefont{F.~J.} \bibnamefont{{Garc\'{\i}a de Abajo}}},
  \bibinfo{author}{\bibnamefont{{A. G. Pattantyus-Abraham}}},
  \bibinfo{author}{\bibfnamefont{N.}~\bibnamefont{Zabala}},
  \bibinfo{author}{\bibfnamefont{A.}~\bibnamefont{Rivacoba}},
  \bibinfo{author}{\bibfnamefont{M.~O.} \bibnamefont{Wolf}}, \bibnamefont{and}
  \bibinfo{author}{\bibfnamefont{P.~M.} \bibnamefont{Echenique}},
  \bibinfo{journal}{Phys.\ Rev.\ Lett.} \textbf{\bibinfo{volume}{91}},
  \bibinfo{pages}{143902} (\bibinfo{year}{2003}).

\bibitem[{\citenamefont{Smith and Purcell}(1953)}]{SP1953}
\bibinfo{author}{\bibfnamefont{S.~J.} \bibnamefont{Smith}} \bibnamefont{and}
  \bibinfo{author}{\bibfnamefont{E.~M.} \bibnamefont{Purcell}},
  \bibinfo{journal}{Phys.\ Rev.} \textbf{\bibinfo{volume}{92}},
  \bibinfo{pages}{1069} (\bibinfo{year}{1953}).

\bibitem[{\citenamefont{Mizuno et~al.}(1987)\citenamefont{Mizuno, Pae,
  Nozokido, and Furuya}}]{MPN1987}
\bibinfo{author}{\bibfnamefont{K.}~\bibnamefont{Mizuno}},
  \bibinfo{author}{\bibfnamefont{J.}~\bibnamefont{Pae}},
  \bibinfo{author}{\bibfnamefont{T.}~\bibnamefont{Nozokido}}, \bibnamefont{and}
  \bibinfo{author}{\bibfnamefont{K.}~\bibnamefont{Furuya}},
  \bibinfo{journal}{Nature} \textbf{\bibinfo{volume}{328}}, \bibinfo{pages}{45}
  (\bibinfo{year}{1987}).

\bibitem[{\citenamefont{Schilling and Raether}(1973)}]{SR1973}
\bibinfo{author}{\bibfnamefont{J.}~\bibnamefont{Schilling}} \bibnamefont{and}
  \bibinfo{author}{\bibfnamefont{H.}~\bibnamefont{Raether}},
  \bibinfo{journal}{J.\ Phys.\ Condens.\ Matter} \textbf{\bibinfo{volume}{6}},
  \bibinfo{pages}{L358} (\bibinfo{year}{1973}).

\bibitem[{\citenamefont{Vesseur et~al.}(2009)\citenamefont{Vesseur,
  {Garc\'{\i}a de Abajo}, and Polman}}]{paper137}
\bibinfo{author}{\bibfnamefont{E.~J.~R.} \bibnamefont{Vesseur}},
  \bibinfo{author}{\bibfnamefont{F.~J.} \bibnamefont{{Garc\'{\i}a de Abajo}}},
  \bibnamefont{and} \bibinfo{author}{\bibfnamefont{A.}~\bibnamefont{Polman}},
  \bibinfo{journal}{Nano\ Lett.} \textbf{\bibinfo{volume}{9}},
  \bibinfo{pages}{3147} (\bibinfo{year}{2009}).

\bibitem[{\citenamefont{{Garc\'{\i}a de Abajo}}(2010)}]{paper149}
\bibinfo{author}{\bibfnamefont{F.~J.} \bibnamefont{{Garc\'{\i}a de Abajo}}},
  \bibinfo{journal}{Rev.\ Mod.\ Phys.} \textbf{\bibinfo{volume}{82}},
  \bibinfo{pages}{209} (\bibinfo{year}{2010}).

\bibitem[{\citenamefont{Barwick et~al.}(2009)\citenamefont{Barwick, Flannigan,
  and Zewail}}]{BFZ09}
\bibinfo{author}{\bibfnamefont{B.}~\bibnamefont{Barwick}},
  \bibinfo{author}{\bibfnamefont{D.~J.} \bibnamefont{Flannigan}},
  \bibnamefont{and} \bibinfo{author}{\bibfnamefont{A.~H.}
  \bibnamefont{Zewail}}, \bibinfo{journal}{Nature}
  \textbf{\bibinfo{volume}{462}}, \bibinfo{pages}{902} (\bibinfo{year}{2009}).

\bibitem[{\citenamefont{Yurtsever and Zewail}(2012)}]{YZ12_2}
\bibinfo{author}{\bibfnamefont{A.}~\bibnamefont{Yurtsever}} \bibnamefont{and}
  \bibinfo{author}{\bibfnamefont{A.~H.} \bibnamefont{Zewail}},
  \bibinfo{journal}{Nano\ Lett.} \textbf{\bibinfo{volume}{12}},
  \bibinfo{pages}{3334} (\bibinfo{year}{2012}).

\bibitem[{\citenamefont{Piazza et~al.}(2015)\citenamefont{Piazza, Lummen,
  {Qui\~{n}onez}, Murooka, Reed, Barwick, and Carbone}}]{PLQ15}
\bibinfo{author}{\bibfnamefont{L.}~\bibnamefont{Piazza}},
  \bibinfo{author}{\bibfnamefont{T.~T.~A.} \bibnamefont{Lummen}},
  \bibinfo{author}{\bibfnamefont{E.}~\bibnamefont{{Qui\~{n}onez}}},
  \bibinfo{author}{\bibfnamefont{Y.}~\bibnamefont{Murooka}},
  \bibinfo{author}{\bibfnamefont{B.}~\bibnamefont{Reed}},
  \bibinfo{author}{\bibfnamefont{B.}~\bibnamefont{Barwick}}, \bibnamefont{and}
  \bibinfo{author}{\bibfnamefont{F.}~\bibnamefont{Carbone}},
  \bibinfo{journal}{Nat.\ Commun.} \textbf{\bibinfo{volume}{6}},
  \bibinfo{pages}{6407} (\bibinfo{year}{2015}).

\bibitem[{\citenamefont{Feist et~al.}(2015)\citenamefont{Feist, Echternkamp,
  Schauss, Yalunin, Sch\"afer, and Ropers}}]{FES15}
\bibinfo{author}{\bibfnamefont{A.}~\bibnamefont{Feist}},
  \bibinfo{author}{\bibfnamefont{K.~E.} \bibnamefont{Echternkamp}},
  \bibinfo{author}{\bibfnamefont{J.}~\bibnamefont{Schauss}},
  \bibinfo{author}{\bibfnamefont{S.~V.} \bibnamefont{Yalunin}},
  \bibinfo{author}{\bibfnamefont{S.}~\bibnamefont{Sch\"afer}},
  \bibnamefont{and} \bibinfo{author}{\bibfnamefont{C.}~\bibnamefont{Ropers}},
  \bibinfo{journal}{Nature} \textbf{\bibinfo{volume}{521}},
  \bibinfo{pages}{200} (\bibinfo{year}{2015}).

\bibitem[{\citenamefont{Photonics and plasmonics in 4D ultrafast~electron
  microscopy}(2015)}]{BZ15}
\bibinfo{author}{\bibnamefont{Photonics}} \bibnamefont{and}
  \bibinfo{author}{\bibnamefont{plasmonics in 4D ultrafast~electron
  microscopy}}, \bibinfo{journal}{ACS\ Photon.} \textbf{\bibinfo{volume}{2}},
  \bibinfo{pages}{1391} (\bibinfo{year}{2015}).

\bibitem[{\citenamefont{{Garc\'{\i}a de Abajo}
  et~al.}(2010)\citenamefont{{Garc\'{\i}a de Abajo}, {Asenjo Garcia}, and
  Kociak}}]{paper151}
\bibinfo{author}{\bibfnamefont{F.~J.} \bibnamefont{{Garc\'{\i}a de Abajo}}},
  \bibinfo{author}{\bibfnamefont{A.}~\bibnamefont{{Asenjo Garcia}}},
  \bibnamefont{and} \bibinfo{author}{\bibfnamefont{M.}~\bibnamefont{Kociak}},
  \bibinfo{journal}{Nano\ Lett.} \textbf{\bibinfo{volume}{10}},
  \bibinfo{pages}{1859} (\bibinfo{year}{2010}).

\bibitem[{\citenamefont{Asenjo-Garcia and {Garc\'{\i}a de
  Abajo}}(2014)}]{paper243}
\bibinfo{author}{\bibfnamefont{A.}~\bibnamefont{Asenjo-Garcia}}
  \bibnamefont{and} \bibinfo{author}{\bibfnamefont{F.~J.}
  \bibnamefont{{Garc\'{\i}a de Abajo}}}, \bibinfo{journal}{Phys.\ Rev.\ Lett.}
  \textbf{\bibinfo{volume}{113}}, \bibinfo{pages}{066102}
  (\bibinfo{year}{2014}).

\bibitem[{\citenamefont{Harvey et~al.}(2015)\citenamefont{Harvey, Pierce,
  Chess, and McMorran}}]{HPC15}
\bibinfo{author}{\bibfnamefont{T.~R.} \bibnamefont{Harvey}},
  \bibinfo{author}{\bibfnamefont{J.~S.} \bibnamefont{Pierce}},
  \bibinfo{author}{\bibfnamefont{J.~J.} \bibnamefont{Chess}}, \bibnamefont{and}
  \bibinfo{author}{\bibfnamefont{B.~J.} \bibnamefont{McMorran}},
  \bibinfo{journal}{arXiv} p. \bibinfo{pages}{1507.0181}
  (\bibinfo{year}{2015}).

\bibitem[{KD1()}]{KD1}
\bibinfo{note}{The present work deals with coherent electron waves. However,
  inelastic energy broadening (sometimes a sizeable fraction of $\hbar\omega$
  in actual experiments) is directly inherited by the transmitted electrons.}

\bibitem[{EPA()}]{EPAPSKD}
\bibinfo{note}{See supplementary material at
  http://link.aps.org/supplemental/xxx.}

\bibitem[{\citenamefont{Jackson}(1999)}]{J99}
\bibinfo{author}{\bibfnamefont{J.~D.} \bibnamefont{Jackson}},
  \emph{\bibinfo{title}{Classical Electrodynamics}}
  (\bibinfo{publisher}{Wiley}, \bibinfo{address}{New York},
  \bibinfo{year}{1999}).

\bibitem[{\citenamefont{{Garc\'{\i}a de Abajo} and
  Manjavacas}(2015)}]{paper254}
\bibinfo{author}{\bibfnamefont{F.~J.} \bibnamefont{{Garc\'{\i}a de Abajo}}}
  \bibnamefont{and}
  \bibinfo{author}{\bibfnamefont{A.}~\bibnamefont{Manjavacas}},
  \bibinfo{journal}{Faraday\ Discuss.} \textbf{\bibinfo{volume}{178}},
  \bibinfo{pages}{87} (\bibinfo{year}{2015}).

\bibitem[{\citenamefont{{Garc\'{\i}a de Abajo}}(2014)}]{paper235}
\bibinfo{author}{\bibfnamefont{F.~J.} \bibnamefont{{Garc\'{\i}a de Abajo}}},
  \bibinfo{journal}{ACS\ Photon.} \textbf{\bibinfo{volume}{1}},
  \bibinfo{pages}{135} (\bibinfo{year}{2014}).

\bibitem[{KD3()}]{KD3}
\bibinfo{note}{A large grating period leads to small diffraction angles, which
  might be difficult to resolve with conventional transmission electron
  microscopes (TEMs). Diffraction from structures with $>100\,$nm period are
  currently resolvable with TEMs \cite{FGP07}. This problem is relaxed for
  low-energy electrons \cite{MPS06}.}

\bibitem[{KD2()}]{KD2}
\bibinfo{note}{T. T. A. Lummen, G. Berruto, R. J. Lamb, L. Dal Negro, F. J.
  Garc\'{\i}a de Abajo, B. Barwick, D. McGrouther, and F. Carbone, to be
  published.}

\bibitem[{\citenamefont{Woessner et~al.}(2015)\citenamefont{Woessner,
  Lundeberg, Gao, Principi, Alonso-Gonz\'alez, Carrega, Watanabe, Taniguchi,
  Vignale, Polini et~al.}}]{WLG15}
\bibinfo{author}{\bibfnamefont{A.}~\bibnamefont{Woessner}},
  \bibinfo{author}{\bibfnamefont{M.~B.} \bibnamefont{Lundeberg}},
  \bibinfo{author}{\bibfnamefont{Y.}~\bibnamefont{Gao}},
  \bibinfo{author}{\bibfnamefont{A.}~\bibnamefont{Principi}},
  \bibinfo{author}{\bibfnamefont{P.}~\bibnamefont{Alonso-Gonz\'alez}},
  \bibinfo{author}{\bibfnamefont{M.}~\bibnamefont{Carrega}},
  \bibinfo{author}{\bibfnamefont{K.}~\bibnamefont{Watanabe}},
  \bibinfo{author}{\bibfnamefont{T.}~\bibnamefont{Taniguchi}},
  \bibinfo{author}{\bibfnamefont{G.}~\bibnamefont{Vignale}},
  \bibinfo{author}{\bibfnamefont{M.}~\bibnamefont{Polini}},
  \bibnamefont{et~al.}, \bibinfo{journal}{Nat.\ Nanotech.}
  \textbf{\bibinfo{volume}{14}}, \bibinfo{pages}{421} (\bibinfo{year}{2015}).

\bibitem[{\citenamefont{Mayer et~al.}(2015)\citenamefont{Mayer, Scarabelli,
  March, Altantzis, Tebbe, Kociak, Bals, {Garc\'{\i}a de Abajo}, Fery, and
  {Liz-Marz\'an}}}]{paper258}
\bibinfo{author}{\bibfnamefont{M.}~\bibnamefont{Mayer}},
  \bibinfo{author}{\bibfnamefont{L.}~\bibnamefont{Scarabelli}},
  \bibinfo{author}{\bibfnamefont{K.}~\bibnamefont{March}},
  \bibinfo{author}{\bibfnamefont{T.}~\bibnamefont{Altantzis}},
  \bibinfo{author}{\bibfnamefont{M.}~\bibnamefont{Tebbe}},
  \bibinfo{author}{\bibfnamefont{M.}~\bibnamefont{Kociak}},
  \bibinfo{author}{\bibfnamefont{S.}~\bibnamefont{Bals}},
  \bibinfo{author}{\bibfnamefont{F.~J.} \bibnamefont{{Garc\'{\i}a de Abajo}}},
  \bibinfo{author}{\bibfnamefont{A.}~\bibnamefont{Fery}}, \bibnamefont{and}
  \bibinfo{author}{\bibfnamefont{L.~M.} \bibnamefont{{Liz-Marz\'an}}},
  \bibinfo{journal}{Nano\ Lett.} \textbf{\bibinfo{volume}{15}},
  \bibinfo{pages}{5427} (\bibinfo{year}{2015}).

\bibitem[{\citenamefont{Lecante et~al.}(1977)\citenamefont{Lecante, Ballu, and
  Newns}}]{LBN1977}
\bibinfo{author}{\bibfnamefont{J.}~\bibnamefont{Lecante}},
  \bibinfo{author}{\bibfnamefont{Y.}~\bibnamefont{Ballu}}, \bibnamefont{and}
  \bibinfo{author}{\bibfnamefont{D.~M.} \bibnamefont{Newns}},
  \bibinfo{journal}{Phys.\ Rev.\ Lett.} \textbf{\bibinfo{volume}{38}},
  \bibinfo{pages}{36} (\bibinfo{year}{1977}).

\bibitem[{\citenamefont{Handali et~al.}(2015)\citenamefont{Handali, Shakya, and
  Barwick}}]{HSB15}
\bibinfo{author}{\bibfnamefont{J.}~\bibnamefont{Handali}},
  \bibinfo{author}{\bibfnamefont{P.}~\bibnamefont{Shakya}}, \bibnamefont{and}
  \bibinfo{author}{\bibfnamefont{B.}~\bibnamefont{Barwick}},
  \bibinfo{journal}{Opt.\ Express} \textbf{\bibinfo{volume}{23}},
  \bibinfo{pages}{5236} (\bibinfo{year}{2015}).

\bibitem[{\citenamefont{Abramowitz and Stegun}(1972)}]{AS1972}
\bibinfo{author}{\bibfnamefont{M.}~\bibnamefont{Abramowitz}} \bibnamefont{and}
  \bibinfo{author}{\bibfnamefont{I.~A.} \bibnamefont{Stegun}},
  \emph{\bibinfo{title}{Handbook of Mathematical Functions}}
  (\bibinfo{publisher}{Dover}, \bibinfo{address}{New York},
  \bibinfo{year}{1972}).

\end{thebibliography}

\end{document}